\documentstyle[12pt,a4]{article}
\begin{document}
\baselineskip = 6.5mm
\topmargin= -15mm
\textheight= 230mm
\begin{center}
\begin{Large}

{\bf  Possible Suppression of Neutron EDM   }

\end{Large}

\vspace{1cm}

Tomoko ASAGA\footnote{JSPS fellow, e-mail: asaga@phys.cst.nihon-u.ac.jp} and
Takehisa FUJITA\footnote{e-mail: fffujita@phys.cst.nihon-u.ac.jp}

Department of Physics, Faculty of Science and Technology

Nihon University, Tokyo, Japan

\vspace{1cm}

{\large ABSTRACT}

\end{center}

Employing generalized Schiff's transformation on electric dipole moments (EDM)
in quantum field theory,
we show that the chromoelectric  EDM lagrangian density
is transformed into the electric EDM term
with a new coefficient.
Under the new constraint on the
EDM operators, the neutron EDM can be described by
a unique combination of electric EDM $d_f$ and chromoelectric EDM
${\tilde d}_f $ of quarks. If the special relation of
$\displaystyle{ d_f={e_f\over{2g_s}}{\tilde d}_f }$ holds,
then the neutron EDM is suppressed significantly.

\vspace{1cm}

\newpage
\begin{enumerate}
\item{\bf Introduction}

Recent measurements of electric dipole moments (EDM) in neutrons and atoms
have presented tight upper bounds [1-5]
which are rather close to theoretical
predictions of the EDM by supersymmetric theory calculations [6-10].
Even though there are some free parameters in the calculations,
it is expected that the finite EDM values may well be observed soon.
In particular, the recent measurement of the muon $g-2$ [23]
 should give severe constraints on the theoretical EDM values.

Recently, there have been many theoretical investigations
of the neutron EDM [11-18].
In particular,  Pospelov and Ritz [19]
 suggest that the contributions to the neutron EDM
which come from the chromoelectric interactions may well be comparable
to the electric EDM estimation. Therefore, careful studies of the
EDM operator arising from the  chromoelectric fields must be quite
important for the reliable estimation of the neutron EDM.

In this Letter, we study a constraint on the chromoelectric  EDM operators
due to the unitary transformation which is analogous to Schiff's theorem in
nonrelativistic quantum mechanics.
Here, we show that the chromoelectric  EDM lagrangian density \ \
$\displaystyle{ -i {{\tilde d_f}\over 2}
\bar \psi \sigma_{\mu \nu}\gamma_5 t^a\psi G^{\mu \nu, a}  }$
is transformed into a new term \ \ $\displaystyle{ i {e_f\over{4g_s}}
{\tilde{d_f}} \bar \psi
\sigma_{\mu \nu}\gamma_5 \psi F^{\mu \nu} }$
which is just the same as the electric EDM lagrangian density. Thus,
 the  coefficient of the EDM lagrangian density is modified
from $\displaystyle{ {d_f\over 2} }$  to
\ $\displaystyle{ {1\over 2}\left( d_f-{e_f\over{2g_s}}\tilde{d_f}\right) }$.
Therefore, any neutron EDM calculations should be described
with the combinations of
\ $\displaystyle{ \left( d_f-{e_f\over{2g_s}}\tilde{d_f}\right) }$.
This should present some constraints which should be satisfied
by any calculations of the neutron EDM.

\vspace{1cm}

\item{\bf Schiff's theorem in QED }

In 1963, Schiff proved [20] that the effect of the EDM interaction cannot be
measured in the nonrelativistic quantum mechanics with electromagnetic
interactions. This can be easily seen
since the hamiltonian with the EDM interaction
$$ H =  { {\bf p}^2\over{2m}} +
e A_0 ( r) -{\bf d}\cdot {\bf E} \eqno{(1)} $$
can be rewritten with the replacement of $\displaystyle{
{\bf r}\rightarrow {\bf r}-{{\bf d}\over e} }$ as
$$ H =  { {\bf p}^2\over{2m}} +
e A_0 ( r) +{\bf d}\cdot \nabla A_0(r) =
 { {\bf p}^2\over{2m}} +
e A_0 (|{\bf r}+{{\bf d}\over e}|)+O(d^2) =
 { {\bf p}^2\over{2m}} +
e A_0 ( r)+O(d^2)
 \eqno{(2)} $$
where ${\bf E}({\bf r}) = -\nabla A_0(r) $, and we have
made use of the fact that the momentum ${\bf p}$ does not
change under the replacement of  $\displaystyle{{\bf r}\rightarrow {\bf r}
-{{\bf d}\over e} }$.
This means that the spectrum of the original hamiltonian with
the EDM term becomes just the same
as that of the hamiltonian without the EDM term since the $O(d^2)$ term
is negligibly small. This is Schiff's theorem, and
the EDM effect cannot be observed in the nonrelativistic quantum mechanics
with electromagnetic interactions.

Next, we treat the field theory version of Schiff's theorem and
 consider  the unitary transformation in QED [21].
The lagrangian density of QED with the electromagnetic EDM terms
can be written as
$$  {\cal L} =
 \bar \psi  i
( \partial_{\mu}+ieA_{\mu})  \gamma^{\mu} \psi  - m_0  \bar \psi  \psi
  -{1\over 4} F_{\mu\nu} F^{\mu\nu}
-i {d_f\over 2} \bar \psi \sigma_{\mu \nu}
 \gamma_5 \psi F^{\mu \nu} \eqno{(3)}  $$
where  $F_{\mu \nu}$ denotes the field tensor  given as
$$ F_{\mu \nu}= \partial_{\mu}A_{\nu}-\partial_{\nu}A_{\mu} .
\eqno{(4)}  $$
Now, we consider the following unitary transformation with
$p_{\mu}=i  \partial_{\mu} $,
$$ \psi' = \exp \left( i{d_f\over e} \gamma_5
p_{\mu}\gamma^{\mu}  \right)
  \psi .  \eqno{(5)} $$
Under this  transformation, the total lagrangian density  becomes,
up to the order of $d_f$
$$ {{\cal L}'} =
 \bar \psi  i
( \partial_{\mu}+ieA_{\mu})  \gamma^{\mu} \psi  - m_0  \bar \psi  \psi
  -{1\over 4} F_{\mu\nu} F^{\mu\nu}
-2i{d_f\over e} \bar \psi \gamma_5 (p_{\mu}p^{\mu} -eA_{\mu}p^{\mu}) \psi .
 \eqno{(6)} $$
This is the new  lagrangian density  in QED and states that
the new EDM term does not couple to the external electric field while
the original EDM term has the coupling with the electric field
${\bf E}$ in the first order as seen in eq.(3).
This is just the same as Schiff's statement that the EDM interaction
does not have the first order perturbation energy with the external
electric field any more.
Here, it should be noted that, even though eq.(6) contains the
$A_{\mu}p^{\mu}$ term, the hamiltonian constructed from eq.(6)
contains only the three dimensional vector potential ${\bf A}\cdot
{\bf p}$ term which does not contribute to the EDM. This is wellknown
in the EDM evaluation in atomic physics [1,21,25].

Therefore, the EDM of a composite system can be generated in the second order
perturbation as
$$ d_n = 2\sum_{N^*} {<N|\sum_i e_iz_i |N^*><N^*|H_{edm}|N>\over{E_N-E_{N^*}} }
\eqno{(7)} $$
where the sum should be taken over excited states in the composite system and
$H_{edm}$ corresponds to the EDM lagrangian term in eq.(6).
$N$ and $N^*$ denote the ground and excited states of the composite system.

At this point, we make a comment on the gauge sysmmetry violation
of the unitary transformation of eq.(5).
This transformation leads to the violation of the gauge
symmetry, but this is justified as long as one is interested in evaluating
the low energy state property of the system up to the order of $g$.
This is the same as Weinberg-Salam treatment of spontaneous symmetry
breaking where the gauge symmetry is broken when evaluating the ground
state property of the system.
Even though Weinberg-Salam's model is not exact, their treatment is
economical. Clearly, one can obtain the same result
with the spontaneous symmetry breaking procedure as the result that is
solved exactly for the low energy property.
But obviously the exact calculation of
the Weinberg-Salam's model is far from economical.

\vspace{1cm}

\item{\bf Unitary transformation in QCD plus QED }

Recently, there have been several papers which treat
the chromoelectric EDM terms [15-19]. The neutron EDM which comes from
the chromoelectric EDM of quarks seems to depend on the model
calculations.  The calculations with QCD sum rules
in ref.[16]
suggest a suppression of the neutron EDM from the chromoelectric EDM
of quarks while the chiral loop estimate gives a sizable contribution
to the neutron EDM [18].
Further,  Pespelov and Ritz [19] show
that the chromoelectric  EDM term gives a significant contribution
to the neutron EDM when they employ QCD sum rules.

In this section, we show that the dominant contributions from the
chromoelectric EDM can be transformed into the identical shape
to the electric EDM term with some modified coefficients.

We start from QCD plus QED lagrangian density with electric and chromoelectric
EDM terms where the charge of the $f$ flavor quark is denoted by $e_f$ ,
$$  {\cal L} =
 \bar \psi  i
( \partial_{\mu}+ig_s A_{\mu}^at^a+ie_fA_{\mu})  \gamma^{\mu} \psi
 - m_0  \bar \psi  \psi
  -{1\over 4} F_{\mu\nu} F^{\mu\nu}
  -{1\over 4} G_{\mu\nu}^a G^{\mu\nu,a} $$
$$-i {{\tilde d}_f\over 2} \bar \psi \sigma_{\mu \nu}
 \gamma_5 t^a\psi G^{\mu \nu, a}
-i {d_f\over 2} \bar \psi \sigma_{\mu \nu}
 \gamma_5 \psi F^{\mu \nu} \eqno{(8)}  $$
where
$$ G_{\mu \nu}^a= \partial_{\mu}A_{\nu}^a-\partial_{\nu}A_{\mu}^a
-g_sC_{abc}A_{\mu}^bA_{\nu}^c. \eqno{(9)}  $$
Now, we consider the following unitary transformation
$$ \psi' = \exp \left( i{{\tilde d}_f\over{2g_s}} \gamma_5
(p_{\mu}-g_sA_{\mu}^at^a) \gamma^{\mu}  \right)
  \psi .  \eqno{(10)} $$
Under this transformation, we obtain the following new lagrangian density,
$$  {\cal L}' =
 \bar \psi  i( \partial_{\mu}+ig_s A_{\mu}^at^a+ie_fA_{\mu})  \gamma^{\mu} \psi
 - m_0  \bar \psi  \psi
  -{1\over 4} F_{\mu\nu} F^{\mu\nu}
  -{1\over 4} G_{\mu\nu}^a G^{\mu\nu,a} $$
$$- {i\over 2}\left(d_f-{e_f\over{2g_s}}{\tilde d}_f\right)
 \bar \psi \sigma_{\mu \nu} \gamma_5 \psi F^{\mu \nu}
-i{ {\tilde d}_f\over{g_s}}  \bar \psi \gamma_5 \left\{
(p_\mu-g_sA_\mu^at^a)^2-e_fA^{\mu} (p_{\mu}-g_sA_{\mu}^a t^a)\right\} \psi .
 \eqno{(11)}  $$
Eq.(11) shows that the chromoelectric EDM term disappears, and instead
the coefficient of the electric EDM term now changes into a new form.
Further, the last term which is similar to the QED case of eq.(6) is generated.
This term gives rise to the neutron EDM in the second order perturbation theory
in the same way as eq.(7).

Here, it should be noted that the two representations of the EDM
operators are indeed equivalent when we estimate their expectation
values with exact wave functions. However, when we consider the EDM of
any composite systems like neutrons, the difference between the two
operators becomes important since we cannot normally obtain
the exact wave functions of neutrons. Employing the new EDM operators,
we can calculate the EDM of neutrons in a transparent and
economical fashion since it can be evaluated in the first order
perturbation theory.

\vspace{1cm}

\item{\bf Neutron EDM }

Now, we want to estimate the neutron EDM based on the lagrangian density
obtained in the previous section. The EDM hamiltonian
which corresponds to the new EDM lagrangian can be written as
$$ H_{edm} = -\sum_f\left\{\left(d_f-{e_f\over{2g_s}}{\tilde
      d}_f\right)
  \gamma_0 {\bf \Sigma \cdot {\bf E}}
  -{i {\tilde d}_f \over{g_s}}\gamma_0 \gamma_5  \left(
    p_{\mu}^2-2 g_s {\bf A}^a\cdot {\bf p}t^a+ g_s^2( A_{\mu}^at^a)^2
\right) \right\} \eqno{(12)}  $$
where we dropped the term which is proportional to the electromagnetic
vector potential $ A_\mu$ since the neutron state
does not contain any photon states in the ground state. Also, the magnetic
term is not included in eq.(12).

Now, we estimate the first term of eq.(12) in the first order perturbation.
Using the SU(6) quark model,
we obtain for the neutron EDM,
$$d_n^{(1)}= \left[ {4\over 3}(d_d-{e_d\over{2g_s}}{\tilde d}_d)
-{1\over 3}(d_u-{e_u\over{2g_s}}{\tilde d}_u) \right]
N_R  \eqno{(13)} $$
where $N_R$ denotes the overlapping integral, and  $N_R=1$ for nonrelativistic
quark models and $N_R \simeq  0.5$ for the MIT
bag model [24].

Next, we evaluate the contribution to the neutron EDM
from  eq.(12) in the second order perturbation as given in eq.(7).
Since there is no ${1\over 2}^- $ state nearby in the neutron excited
states, there is no enhancement, contrary to the atomic cases [21].

Here, we carry out a naive estimation of eq.(7)
with the MIT bag model [24], and therefore, we  employ the closure
approximation. In this case, eq.(7) becomes
$$ d_n^{(2)}={2\over{\Delta E}}<N| \sum_{i=d,d,u} e_iz_i
 \sum_{j=d,d,u} \left\{  \left(d_j-{e_j\over{2g_s}}{\tilde d}_j\right)
\gamma_0^{(j)} {\bf \Sigma}^{(j)} \cdot {\bf E}^{(j)}
\right.
 $$
$$-  \left. {i{\tilde d}_j\over{g_s}}\gamma_0^{(j)}\gamma_5^{(j)}
\left( {p_{\mu}^{(j)}}^2-2 g_s {\bf A}^a\cdot {\bf p}^{(j)}t_j^a
  + g_s^2( A_{\mu}^at_j^a)^2
\right) \right\} |N> \eqno{(14)} $$
where $\Delta E$ is an average excitation energy of neutron ${1\over 2}^-$
states which should be of the order of nucleon mass $M_N$. ${\bf E}^{(j)}$
denotes the electric field which the $j-$th quark feels inside the bag.
After some numerical calculations, we obtain
$$ d_n^{(2)} \simeq {2\over{3R_0\Delta E}} \left[ 2e_d^2
\left(d_d-{e_d\over{2g_s}}{\tilde d}_d\right)+e_u^2
\left(d_u-{e_u\over{2g_s}}{\tilde d}_u\right)\right] +
{4R_0\over{9g_s\Delta E}} \left( 2e_dm_d^2{\tilde d}_d +
 e_um_u^2{\tilde d}_u \right)
 \eqno{(15)}  $$
where $m_d$ ($m_u$)
denotes the   mass of the $d$ ($u$) quarks. Now,
we take the $\Delta E \sim M_N$ and $R_0\sim
{1\over{m_\pi}}$ with  $m_{\pi}$  the pion mass, and
therefore, the total neutron EDM can be written as
$$d_n= \left[ {4\over 3}\left(d_d-{e_d\over{2g_s}}{\tilde d}_d\right)
-{1\over 3}\left(d_u-{e_u\over{2g_s}}{\tilde d}_u\right) \right] N_R $$
$$+{2m_\pi\over{3M_N}} \left[ 2e_d^2
\left(d_d-{e_d\over{2g_s}}{\tilde d}_d\right)+e_u^2
\left(d_u-{e_u\over{2g_s}}{\tilde d}_u\right)\right]
+ {4\over{9g_s m_{\pi}M_N}}
\left( 2e_dm_d^2{\tilde d}_d + e_um_u^2{\tilde d}_u \right) . \eqno{(16)}  $$
Since the second line of eq.(16) is much smaller than the first line
contribution,
we can practically neglect the second order perturbation contribution
to the neutron EDM
 as long as the first two terms (the first line) survive.

\vspace{1cm}

\item{\bf Possible suppression of neutron EDM }

As can be seen from eq.(16), the neutron EDM must be significantly suppressed
if the following relation holds for $f$ flavor quark,
$$ d_f={e_f\over{2g_s}}{\tilde d}_f  . \eqno{(17)}  $$
Recently, Chang, Keung and Pilaftsis [22] presented the two-loop calculation
to the EDM in supersymmetric theories.
They show that the electric EDM $d_f$ and the chromoelectric
EDM ${\tilde d}_f$ of a light fermion $f$ can be given as
$$ d_f = e_f {3\alpha_{em}\over{64\pi^3}} {R_fm_f\over{M_a^2}}
\sum_{q=t,b}Q_q^2 H_q \eqno{(18a)} $$
$$ {\tilde d}_f = g_s {\alpha_{s}\over{128\pi^3}} {R_fm_f\over{M_a^2}}
\sum_{q=t,b} H_q \eqno{(18b)} $$
where $\alpha_{em}$ and $\alpha_{s}$ are the electromagnetic
 and the chromomagnetic fine structure constants, respectively.
Also, $R_f$ is given as $R_f=\cot \beta $,
 and $M_a$ denotes the tree level mass of the CP-odd Higgs.
 $H_q$ is defined as
$$ H_q= \xi_q\left[ F\left({M_{ \tilde{q}_1}^2\over{M_a^2}}\right)
- F\left({M_{ \tilde{q}_2}^2\over{M_a^2}}\right) \right]   \eqno{(19)} $$
where $F(z)$ denotes a two loop function as given in ref.[22],
$$ F(z)= \int_0^1dx {x(1-x)\over{z-x(1-x)}} \ln\left[ {x(1-x)\over{z}}
\right] . $$
Also, $\xi_q$'s are the CP violating couplings
which are given in the MSSM scheme [7].

Now, we want to see the relation between  $ d_f$ and  ${\tilde d}_f$.
From eqs.(18), we find
$$ d_f={e_f\over{2g_s}}{\tilde d}_f  \left\{ \left(
{12\alpha_{em}\over{\alpha_s}}\right)
 {\sum_{q=t,b} Q_q^2 H_q
\over{\sum_{q=t,b} H_q}}  \right\} . \eqno{(20)}  $$
Since the value of
$\displaystyle{ \left( {12\alpha_{em}\over{\alpha_s}}\right) }$ is
close to unity at the electroweak scale, the parameters in $H_q$
will control the magnitude of the curly bracket of eq.(20).
If the value of the curly bracket is  unity, then
the relation of eq.(17) holds.

At the present stage, we do not know whether the parameters that appear
in eqs.(18) can be estimated reliably or not in the supersymmetric
theories. If these parameters can be determined reliably, it would be
very interesting to examine  the validity of the relation.

\vspace{1cm}

We thank T. Nihei and M. Hiramoto
for helpful discussions and comments.
This work is supported in part by
Japan Society for the Promotion of Science.

\vspace{3cm}
\underline{\bf References}
\vspace{0.5cm}

1. I.B.Khriplovich and S.K.Lamoreaux, {\it CP Violation Without Strangeness},

\qquad Springer 1997

2. K.F. Smith $et$ $al.$, Phys. Lett. {\bf B234} (1990) 191

3. P.G. Harris  $et$ $al.$, Phys. Rev. Lett.  {\bf 82} (1999) 904

4. E.D. Commins $et$ $al.$, Phys. Rev. {\bf A50} (1994) 2960

5. M.V. Romalis, W.C. Griffith, and E.N. Fortson,
Phys. Rev. Lett.  {\bf 86} (2001) 2505

6. S.M. Barr and A. Zee, Phys. Rev. Lett.  {\bf 65} (1990) 21

7. A. Pilaftsis, Phys. Lett.  {\bf B435} (1998) 88,

\qquad Phys. Rev.  {\bf D58} (1998) 096010

8. H. Georgi and S. Dimopoulos, Nucl. Phys. {\bf B193} (1981) 150

9. N. Sakai, Z. Phys. {\bf C11} (1981) 153

10. A.G. Cohen, D.B. Kaplan, and A.E. Nelson, Phys. Lett.  {\bf B388} (1996) 588

11.  H. Georgi and A. Manohar, Nucl. Phys. {\bf B234} (1984) 189

12. R. Arnowitt, J.L. Lopez, and D.V. Nanopoulos,  Phys. Rev. {\bf D42} (1990) 2423

13. R. Arnowitt, M.J. Duff, and K.S. Stelle,  Phys. Rev. {\bf D43} (1991) 3085

14. T. Ibrahim and P. Nath, Phys. Rev. {\bf D57} (1998) 478

15. M. Pospelov and P. Nath,  Phys. Rev. Lett.  {\bf 38} (1999) 2526

16. V.M. Khatsimovsky, I.B.Khriplovich, and A.S. Yelkhovsky,

\qquad Ann. Phys. {\bf 186} (1988) 1

17. L.L. Kogan and D. Wyler,  Phys. Lett. {\bf B274} (1992) 100

18. V.M. Khatsimovsky and  I.B.Khriplovich,  Phys. Lett. {\bf B296} (1994) 219

19.  M. Pospelov and A. Ritz, Phys. Rev {\bf D63} (2001) 073015

20. L.I.Schiff, Phys. Rev. {\bf 132} (1963) 2194

21. T. Asaga, T. Fujita and M. Hiramoto, Prog. Theor. Phys. {\bf 106} (2001) 1223

22. D. Chang, W.Y. Keung, and A.  Pilaftsis, Phys. Rev. Lett.
 {\bf 82} (1999) 900

23. H.N. Brown  $et$ $al.$, Muon ($g-2$) Collaboration, hep-ex/0102017

24. A. Chodos  $et$ $al.$,  Phys. Rev. {\bf D9} (1974) 3471

25. W.R.Johnson, D. S.Guo, M. Idrees and J. Sapirstein,

\qquad Phys. Rev. {\bf A34} (1986) 1043

\end{enumerate}

\end{document}